\documentclass[10pt,a4paper]{article}
\usepackage{geometry}
\usepackage{graphicx}
\usepackage{subfig}
\usepackage{cite}
\usepackage{amsmath}

\begin{document}

\title{Novel layout for reflective arrayed waveguide gratings based on Sagnac loop reflectors with custom spectral response}

\author{Bernardo~Gargallo~~and~Pascual~Mu\~noz\footnote{B. Gargallo and P. Munoz are with the Optical and Quantum Communications Group, iTEAM Research Institute, Universitat Polit\`ecnica de Val\`encia, C/ Camino de Vera s/n, Valencia 46022, Spain. P. Munoz is as well with VLC Photonics S.L., C/ Camino de Vera s/n, Valencia 46022, Spain. bergarja@iteam.upv.es}}

\maketitle


\begin{abstract}
A novel layout for a reflective Arrayed Waveguide Grating is presented in this paper. The device consists of one half of a regular AWG where each arm waveguide in the array is terminated with a phase shifter and a Sagnac loop reflector. The theoretical model to describe and design the spectral response of this device is derivated. By individually adjusting the phase shifter and Sagnac reflectivity in each arm, this layout allows for tailoring the spectral response of the AWG. Design cases of Silicon-on-Insulator AWG devices with Gaussian and flattened response are provided.
\end{abstract}


\section{Introduction}

Wavelength multi/demultiplexers are central components in optical telecommunication networks, and they have been subject of intense research and development since the advent of wavelength-division multiplexing (WDM) in the early 90s\cite{brackett}. Components for these networks are subject to demanding requirements, both in terms of performance and manufacturing. While performance depends on the particular component, all need to provide stable operation. In terms of manufacturing requirements, a reproducible and mass fabrication process is mandatory. Hence, photonic integration is usually the basis for large count WDM multiplexers. The cost of an integrated circuit is fundamentally related to its footprint \cite{kirchain_nphot, pennings_jstqe1996}, and it has a direct impact on the economies of scale for manufacturing, where in general more devices per wafer, i.e. more compact devices, are desired. 

Amongst the different implementations for multiplexers, the Arrayed Waveguide Grating (AWG) \cite{smit, dragone} is one of a few that aligns with the previous statements. Traditionally manufactured on Silica on Silicon integration technology \cite{dragone_ptl1991}, it finds room nearly in all the relevant material platforms, such as Indium Phosphide, Silicon on Insulator, Silicon and Silicon Nitride (for a summary see \cite{munoz_icton2013}). The physical layout of an AWG consists of the combination of waveguides and slab couplers \cite{smit, dragone}. In its most common shape, two slab couplers with input/output waveguides are inter-connected by a set of waveguides, usually referred as arrayed waveguides (AWs). Consecutive waveguides in the array have a length differing a constant amount, which imposes a wavelength-dependent lineal phase front on the signal fed from the first slab coupler. This linear phase front, in combination with the second slab coupler, enable the spatial separation of different wavelengths in different outputs.

In terms of footprint, other integrated multiplexer implementations as the Echelle Diffraction Grating (EDG) achieve considerable size reduction compared to the AWG \cite{lycett}. The layout of an EDG includes a single slab coupler, with input/output waveguides on one side, and a reflective grating on the opposite end. It is a so-called reflective multi/demultiplexer. One issue with EDGs is to maximize the reflection on the grating, in order to minimize the overall insertion losses, issue which is otherwise not present in a regular AWG. Different approaches exist to increase the reflectivity of the grating in an EDG, the most employed being the deposition of metal layers at the edge of the grating \cite{feng}, or the addition to the grating of other structures such as Bragg reflectors \cite{ryckeboer}.  While the former supplies broadband reflectors, it requires resorting to additional fabrication steps. Conversely, Bragg reflectors can be manufactured in the same steps that the EDG, but it is well known the reflection bandwidth is inversely proportional to their strength \cite{pruessner}.

Similarly, AWG layouts with reflective structures midway in the array, i.e. reflective AWGs (R-AWG) are possible as well. They can have a footprint ideally half of a regular AWG, and closer to the one of an EDG. Hence, the signals travelling in the arrayed waveguides arrive to the reflectors, and are bounced back to the (single) slab coupler. Although the functionality is the same than in the case of a regular AWG, some additional design considerations are required \cite{peralta}. The reflectors can be implemented in similar ways to the ones for the EDGs, and the literature shows solutions as reflective coatings on a fact of the chip where the arrayed waveguides end \cite{inoue, soole}, photonic crystals \cite{dai}, external reflectors \cite{peralta2} and even Bragg reflectors \cite{okamoto_dbr} at the end of the arrayed waveguides.

A common issue of all the described approaches for the reflector is that broadband full reflectivity requires additional fabrication steps, and therefore increases the final cost of the multiplexer. In this paper we propose a novel configuration for a R-AWG, where the well known Sagnac Loop Reflectors (SLR) are used as reflective elements at the end of the arrayed waveguides. A SLR is composed of an optical coupler with two output waveguides, that are connected to each other forming a loop. These reflectors are broadband, can supply total reflection, and can be fabricated in the same lithographic process than the rest of the AWG. Moreover, the reflection of a SLR depends on the coupling constant of the coupler. Hence, it can be different for each of the waveguides in the array. The modification of the field pattern in the arrayed waveguides of an AWG allows for spectral response shaping, as for example box like transfer function \cite{okamoto} and multi-channel coherent operations \cite{doerr} amongst other.

The paper is structured as follows. In Section \ref{sec:model}, the theoretical equations describing the full field (amplitude and phase) transfer function of the R-AWG are developed, following the model in \cite{pascual}. The equations are then particularized for two cases: firstly, the case for which all the SLRs have total reflection, and the AWG response obtained is Gaussian; secondly, the case where the reflectivity of each SLR is designed to be such a way that the field profile in the array is a sinc function as in \cite{okamoto} to obtain a flat-top response at the output. In Section \ref{sec:sims}, the equations are used to design and simulate a R-AWG on Silicon-on-Insulator (SOI) technology. Finally, Section \ref{sec:conclusion} presents the summary and outlook.

\begin{figure}
	\centering
	\includegraphics[width=0.8\textwidth]{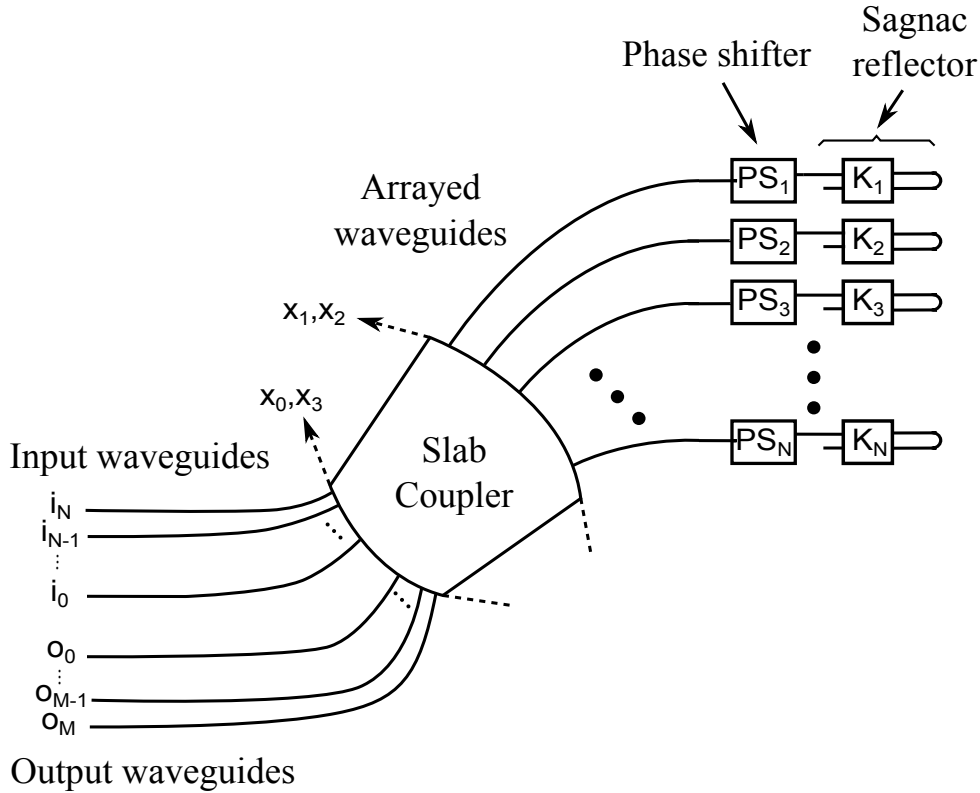}
	\caption{R-AWG schematic view. Abbreviations: $PS$ phase shifter, $K$ coupling constant, $x_i$ (i=0,1,2,3) are reference coordinates and $i_{j}$, $o_{j}$ are input and output waveguides, respectively.}
	\label{fig:srawg}
\end{figure}

\section{\label{sec:model}R-AWG theoretical model}
The schematic view for a reflective AWG (R-AWG) shown in Fig.~\ref{fig:srawg} is used as reference for the equations in this section. As a regular AWG, the layout consists of a group of input and output waveguides connected to one side of the (single) slab coupler. Each of the arrayed waveguides, which are connected to the opposite side of the slab coupler, is terminated with a SLR. The lengths of consecutive waveguides in array differ by a constant amount \cite{smit}. The layout in Fig.~\ref{fig:srawg} includes a phase shifter (PS) section in between the waveguides in the array and the SLR, the purpose of which will be detailed later on.

Hence, this configuration allows for adjusting independently the field amplitude and phase in each AW. Though the operation is similar to a regular AWG, is summarized here for completeness. The field introduced through an input waveguide is diffracted in the slab coupler, and collected by the arrayed waveguides on the opposite side. Then the light travels forth and back through each individual and independent AW, PS and SLR. The reflection (amplitude and phase) in each SLR can be different, depending on the coupling constant for the coupler. The overall reflected field reaching back the slab coupler will be diffracted by the AWs. The overall phase relations between AWs will determine the R-AWG behaviour. In the most simple case, a constant phase difference between consecutive propagation paths in the array will spatially separate the different wavelengths on the input/output side of the slab coupler.

\subsection{Elements}
\begin{figure}
	\centering
	\subfloat[] {
		\label{fig:sagnac_1coupler}
		\includegraphics[width=0.35\textwidth]{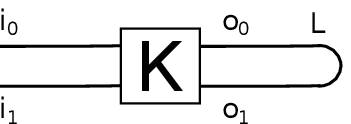}}
		\hspace{0.1\textwidth}
	\subfloat[] {
		\label{fig:sagnac_2couplers}
		\includegraphics[width=0.35\textwidth]{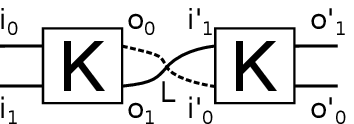}}
	\caption{Sagnac Loop Reflector (a) and SLR analysis as two serial couplers (b). Abbreviations: $i$ and $o$ stand for input and output waveguides, respectively. $K$ stands for coupling constant and  $L$ stands for loop length.}
\label{fig:sagnac}
\end{figure}

Though well known, further details are given for the SLR, for which a reference layout is shown in Fig.~\ref{fig:sagnac_1coupler}. 

The transfer matrix for the SLR can be expressed as:
\begin{equation}
	\begin{pmatrix} o_{0} \\ o_{1} \end{pmatrix} = \begin{pmatrix} \sqrt{1-k}  & j\sqrt{k} \\ j\sqrt{k}  & \sqrt{1-k} \end{pmatrix} \begin{pmatrix} i_{0} \\ i_{1} \end{pmatrix}
\end{equation}

Hence, it is possible to analyze the SLR as two couplers connected in a series, Fig.~\ref{fig:sagnac_2couplers}. For the case where only the input $i_{0}$ is used, the transfer functions are:
\begin{equation}
	o'_{0} = 2j\sqrt{\left( 1-k \right) k} e^{-j\beta L}i_{0}
\end{equation}
\begin{equation}
	o'_{1} = \left( 1-2k \right) e^{-j \beta L}i_{0}
\end{equation}
which at the reproduced the well known total reflection to $i_{0}$ when the coupling constant is set to $k = 0.5$. Note that the equations include a phase change due to the length of the loop, $L$. For coupling constants $k$ other than 0.5, the reflected power will be less than 100\%.

The coupler for the SLR can be implemented in multiples ways: directional coupler (DC) \cite{saleh}, wavelength insensitive coupler (WINC) \cite{jinguji} or multimode interference coupler (MMI) \cite{soldano}. The reflectors for the proposed R-AWG need to be broadband, i.e. the coupling constant needs to be constant over a wide range of wavelengths. Therefore only the WINC and MMI couplers satisfy this condition. Moreover, footprint considerations lead to the selection of MMI vs WINC, since the latter is in general larger than the former. Finally, MMIs can be designed to have arbitrary coupling constant, as described in \cite{bachmann, besse}. Different coupling constants may ultimately result in different MMI lengths/phase shifts.

The purpose of the formerly introduced PS sections is to compensate the phase imbalances between AWs due to different phase shifts/coupling constant/reflection between the SLRs of the different AWs. As the coupler, the phase shifter is required to be broadband. This is possible by means of regular straight waveguides, or by tapered waveguides as in \cite{jeong}. 

\subsection{Principle of operation}
The principle of operation for the AWG requires that the phase shift ($ \Delta \phi$) between two consecutive AWs is integer number ($m$) times of $2\pi$.
In the most general case of Fig.~\ref{fig:srawg}, the total phase shift the light undergoes in each AW can be given by:
\begin{equation}
	\phi_{i} = \phi_{w,i} + \phi_{PS,i} + \phi_{SLR,i}
\end{equation}
where $i$ is the number of the waveguide and subscripts $w$, $PS$ and $SLR$ stand for waveguide, Phase Shifter and Sagnac Loop Reflector, respectively. Consider the field in an input waveguide, placed for simplicity at the center input side of the slab coupler, and approximated by the following normalized Gaussian function:
\begin{equation}
	b_{i} \left( x_{0} \right) = \sqrt[4]{\frac{2}{\pi \omega_{i}^{2}}} e^{- \left( \frac{x_{0}}{\omega_{i}} \right)^{2}}
\end{equation}
being $\omega_{i}$ the mode field radius and $x_{0}$ the spatial coordinate at the input plane. This field is radiated to the AW side of the slab coupler, where the light spatial distribution can be obtained by the spatial Fourier transform of the input profile, using the paraxial approximation \cite{goodman}:
\begin{equation}
	B_{i} \left( x_{1} \right) = \mathcal{F} \left. \left\lbrace b_{i} \left( x_{0} \right) \right\rbrace \right|_{u=\frac{x_{1}}{\alpha}} = \sqrt[4]{2\pi \frac{\omega_{i}^{2}}{\alpha^{2}}} e^{- \left( \pi \omega_{i} \left( \frac{x_{1}}{\alpha} \right) \right)^{2}}
\end{equation}
where $u$ is the spatial frequency domain variable of the Fourier transform and $\alpha$ is the equivalent to the wavelength focal length product in Fourier optics propagation, expressed as:
\begin{equation}
	\alpha = \frac{cL_{f}}{n_{s}\nu}
\end{equation}
being $L_{f}$ the slab length, $n_{s}$ the effective index of the slab coupler mode, $\nu$ the frequency and $c$ the speed of light in vacuum. The total field distribution for an arbitrary number of this way illuminated AWs, placed at the $x_{1}$ plane, is:
\begin{equation}
	f_{1} \left( x_{1} \right) = \sqrt[4]{2\pi \omega_{g}^{2}} \sum_{r} B_{i} \left( r d_{\omega} \right) b_{g} \left( x_{1}-rd_{\omega} \right)
\end{equation}
where $r$ is the AW number, $d_{\omega}$ is the spacing, $\omega_{g}$ is the mode field radius and $b_{g} \left( x \right)$ is the field profile of the AWs. Particularizing for a set of an arbitrary number $N$ waveguides:
\begin{equation} \label{eq:f1}
	f_{1} \left( x_{1} \right) = \sqrt[4]{2\pi \omega_{g}^2} \left[ \prod \left( \frac{x_{1}}{Nd_{\omega}} \right) B_{i} \left( x_{1} \right) \sum_{r=-\infty}^{+\infty} \delta \left( x_{1} - rd_{\omega} \right) \right] \otimes  b_{g} \left( x_{1} \right)
\end{equation}
with $\otimes$ being the convolution, and $\prod \left( x_{1} / Nd_{\omega} \right)$ being a truncation function defined as:
\begin{equation}
 \prod \left( \frac{x_{1}}{Nd_{\omega}} \right) = \left\{ \begin{array}{ll}
        1, & \mbox{if $\left| x \right| \leq \frac{Nd_{\omega}}{2}$}\\
        0, & \mbox{otherwise} \end{array} \right. 
\end{equation}
Let the length of waveguide number $r$ be given by:
\begin{equation}
	l_{r} = \frac{l_{0}}{2} + \frac{\Delta l}{2} \left( r + \frac{N}{2} \right)
\end{equation}
where $l_{0}$/2 is the (base) length of the shortest waveguide. The value of the incremental length between arrayed waveguides ($\Delta l$/2) is set to an integer multiple $m$, known as AWG grating order, of the named central design wavelength $\lambda_{0}$:
\begin{equation}
	\Delta l = \frac{m\lambda_{0}}{n_{c}}
\end{equation}
where $n_{c}$ is the effective index in the waveguides. The value of $\Delta l$ ensures that the lightwave from a central input waveguide will be focused on a central output waveguide in a regular AWG for $\lambda_0$. The phase shift introduced by waveguide $r$ will be:
\begin{equation}
	\Delta \phi_{r} = \beta l_{r} = 2 \pi \frac{n_{c}}{c} \nu l_{r}
\end{equation}
where $\beta$ is the propagation constant of the (single) mode in the waveguide. The latter term is introduced in Eq. (\ref{eq:f1}) to obtain:
\begin{equation} \label{eq:f1p}
	f'_{1} \left( x_{1}, \nu \right) = \sqrt[4]{2\pi \omega_{g}^2} \left[ \prod \left( \frac{x_{1}}{Nd_{\omega}} \right) B_{i} \left( x_{1} \right) \phi \left( x_{1}, \nu \right) \sum_{r=-\infty}^{+\infty} \delta \left( x_{1} - rd_{\omega} \right)  \right] \otimes  b_{g} \left( x_{1} \right) 
\end{equation}
where $\phi \left( x_{1}, \nu \right)$ is defined as:
\begin{equation}
	\phi \left( x_{1}, \nu \right) = \psi \left( \nu \right) e^{-j\pi m \frac{\nu}{\nu_{0}} \frac{x_{1}}{d_{\omega}}}
\end{equation}
\begin{equation}
	\psi \left( \nu \right) = e^{-j 2 \pi \nu \left( \frac{n_{c} l_{0}}{2c} + \frac{mN}{4\nu_{0}} \right)}
\end{equation}
Eq. (\ref{eq:f1p}) shows the field at the input of the phase shifters. Both the PS and the SLR will introduce an additional phase shift, and the SLR an amplitude change, hence:
\begin{equation}
\begin{aligned}
	f''_{1} \left( x_{1}, \nu \right) = \sqrt[4]{2\pi \omega_{g}^2} \Biggl[ \prod \left( \frac{x_{1}}{Nd_{\omega}} \right)  & B_{i} \left( x_{1} \right) \phi \left( x_{1}, \nu \right) \\
	 & \sum_{r=-\infty}^{+\infty} \delta \left( x_{1} - rd_{\omega} \right) e^{-j \psi_{PS,r} \left( \nu \right)} j A_{r} e^{-j\beta l_{SLR,r}} \Biggr] \otimes b_{g} \left( x_{1} \right) 
\end{aligned}
\end{equation}
where $A_{r}$ is the SLRs amplitude term given by:
\begin{equation}
	A_{r} = 2 \sqrt{\left( 1-k_{r} \right) k_{r}}
\end{equation}
and $\psi_{PS,r} \left( \nu \right)$ is the phase shift introduced by the PS, $k_{r}$ is the SLR coupling constant and $l_{SLR,r}$ is the length of the loop waveguide within the SLR. The reflected field from the SLRs at the plane $x_{2}$, which is the same that the plane $x_{1}$ in a R-AWG, is given by:
\begin{equation} \label{eq:f2}
\begin{aligned}
	f_{2} \left( x_{2}, \nu \right) = \sqrt[4]{2\pi \omega_{g}^2} \Biggl[ \prod \left( \frac{x_{2}}{Nd_{\omega}} \right) &B_{i} \left( x_{2} \right) \phi' \left( x_{2}, \nu \right)  \\
	&\sum_{r=-\infty}^{+\infty} \delta \left( x_{2} - rd_{\omega} \right) e^{-2j \psi_{PS,r} \left( \nu \right)} j A_{r}  e^{-j\beta l_{SLR,r}} \Biggr] \otimes b_{g} \left( x_{2} \right) 
\end{aligned}
\end{equation}
where now the phase term $\phi' \left( x_{2}, \nu \right)$ is:
\begin{equation}
	\phi' \left( x_{2}, \nu \right) = \psi' \left( \nu \right) e^{-j2\pi m \frac{\nu}{\nu_{0}} \frac{x_{2}}{d_{\omega}}}
\end{equation}
\begin{equation}
	\psi' \left( \nu \right) = e^{-j 2 \pi \nu \left( \frac{n_{c} l_{0}}{c} + \frac{mN}{2\nu_{0}} \right)}
\end{equation}
The field at the plane $x_{3}$ (that is the same that $x_{0}$ in a R-AWG) can be calculated using the spatial Fourier transform as:
\begin{equation} \label{eq:f3}
	f_{3} \left( x_{3}, \nu \right) = \mathcal{F} \left. \left\lbrace f_{2} \left( x_{2}, \nu \right) \right\rbrace \right|_{u=\frac{x_{3}}{\alpha}}
\end{equation}
Contrary to our previous model in \cite{pascual} where a closed analytical solution for the field at the output plane is derived, no straightforward closed analytical solution is possible in the general case, due to the arbitrary phase shift for each AW. Nonetheless, the previous equation is the basis for the particular cases derived in the next paragraphs. Independently, the frequency response at the output waveguide $q$ can be calculated through the following overlap integral:
\begin{equation}
\label{eq:tq}
	t_{q} \left( x_{3} \right) = \int_{-\infty}^{+\infty} f_{3} \left( x_{3},\nu \right) b_{0} \left( x_{3}- qd_{o} \right) \partial x_{3}
\end{equation}
where $d_{o}$ is the spacing between, and $b_0\left(x_3\right)$ is the field profile of, the output waveguides.

\subsubsection{\label{subsec:gaussian}Gaussian spectral response}
The basic case for the R-AWG concept introduced in this paper is when all the SLRs are equal and with total reflection, i.e. coupling constant $k=0.5$. Since the SLRs are ideally identical, no phase shifters are required in this configuration. The layout for this configuration is the same than in Fig.~\ref{fig:srawg}, where the phase shifters have been removed, all the SLRs are identical and therefore the length between consecutive AW differ by an incremental length $\Delta l/2$.
The equations can be particularized to this case, i.e. Eq. (\ref{eq:f2}) can be rewritten as:
\begin{equation} \label{eq:f2_sinsinc}
	f_{2} \left( x_{2}, \nu \right) = j \sqrt[4]{2\pi \omega_{g}^2} e^{-j \beta l_{SLR}} \Biggl[ \left. \prod \left( \frac{x_{2}}{Nd_{\omega}} \right) B_{i} \left( x_{2} \right) \phi' \left( x_{2}, \nu \right) \sum_{r=-\infty}^{+\infty} \delta \left( x_{2} - rd_{\omega} \right) \right] \otimes b_{g} \left( x_{2} \right) 
\end{equation}
and the field at the output plane ($x_{3}$) described using Eq. (\ref{eq:f3}) as:
\begin{equation}
\begin{aligned}
	f_{3} \left( x_{3}, \nu \right) = j \sqrt[4]{2\pi \frac{\omega_{g}^{2}}{\alpha^{2}}} B_{g} \left( x_{3} \right) e^{-j \beta l_{SLR}} \Biggl[ b_{i} \left( x_{3} \right) &\otimes \mathrm{sinc} \left( N d_{\omega} \frac{x_{3}}{\alpha} \right) \\
	 & \otimes \Phi \left( x_{3}, \nu \right) \otimes \sum_{r=-\infty}^{+\infty} \delta \left( x_{3} -r\frac{\alpha}{d_{\omega}} \right) \Biggr]
\end{aligned}
\end{equation}
where the different terms therein are given by:
\begin{equation}
b_{i} \left( x_{3} \right) = \mathcal{F} \left. \left\lbrace B_{i} \left( x_{2} \right) \right\rbrace \right|_{u=\frac{x_{3}}{\alpha}}
\end{equation}
\begin{equation}
B_{g} \left( x_{3} \right) = \mathcal{F} \left. \left\lbrace b_{g} \left( x_{2} \right) \right\rbrace \right|_{u=\frac{x_{3}}{\alpha}}
\end{equation}
\begin{equation}
	\Phi \left( x_{3}, \nu \right) = \mathcal{F} \left. \left\lbrace \phi' \left( x_{2}, \nu \right) \right\rbrace \right|_{u=\frac{x_{3}}{\alpha}} = \phi' \left( \nu \right) \delta \left( x_{3} + \frac{\alpha m}{d_{\omega} \nu_{0}} \nu \right)
\end{equation}
and therefore,
\begin{equation} \label{eq:f3_sinsinc}
	f_{3} \left( x_{3}, \nu \right) = j \sqrt[4]{2\pi \frac{\omega_{g}^{2}}{\alpha^{2}}} B_{g} \left( x_{3} \right) e^{-j \beta l_{SLR}} \psi' \left( \nu \right) \sum_{r=-\infty}^{+\infty} f_{M} \left( x_{3} -r\frac{\alpha}{d_{\omega}} + \frac{\nu}{\gamma} \right)
\end{equation}
where $f_{M} \left( x_{3} \right)$ is defined as:
\begin{equation}
	f_{M} \left( x_{3} \right) = \mathrm{sinc} \left( N d_{\omega} \frac{x_{3}}{\alpha} \right) \otimes b_{i} \left( x_{3} \right)
\end{equation}
Recall during the first steps in the derivation of the model we assumed for simplicity the input waveguide is centered with respect to the slab coupler. Nonetheless Eq. (\ref{eq:f3_sinsinc}) can be readily extended to account for the other input positions, i.e. for a the field at input waveguide number $p$, the equation can be then expressed as follows:
\begin{equation} \label{eq:bip}
	b_{i,p} \left( x_{0} \right) = \sqrt[4]{\frac{2}{\pi \omega_{i}^{2}}} e^{- \left( \frac{x_{0}-pd_{i}}{\omega_{i}} \right)^{2}} = b_{i} \left( x_{0} - pd_{i} \right)
\end{equation}
where $d_{i}$ is the distance between the input waveguides. Then, rewriting Eq. (\ref{eq:f3_sinsinc}) using (\ref{eq:bip}):
\begin{equation} \label{eq:f3_p}
	f_{3,p} \left( x_{3}, \nu \right) = j \sqrt[4]{2\pi \frac{\omega_{g}^{2}}{\alpha^{2}}} B_{g} \left( x_{3} \right) e^{-j \beta l_{SLR}} \psi' \left( \nu \right)	\sum_{r=-\infty}^{+\infty} f_{M} \left( x_{3} + pd_{i}-r\frac{\alpha}{d_{\omega}} + \frac{\nu}{\gamma} \right)
\end{equation}
In conclusion, the functionality of the R-AWG and AWG can be described by similar formulation \cite{pascual}. One important difference in the case of a R-AWG is the positioning of the input/output waveguides, which has implications for the selection of the central design wavelength, however this can be readily accounted for during design as described in \cite{peralta}. The dispersion angle ($\theta$) with respect to the center of the slab coupler is given by \cite{smit}:
\begin{equation}
\label{eq:theta}
	\theta = \arcsin \left( \frac{\beta \Delta l - m2\pi}{\beta_{S} d_{\omega}} \right)
\end{equation}
where $\beta$ and $\beta_{S}$ are the propagation constants of the AW mode and slab modes, respectively, and $d_{\omega}$ is the spacing between AWs. 
For the positioning of the input/output waveguides in the R-AWG, the wavelength routing properties of the AWG need to be observed \cite{takahashi}. Let $\lambda_{p,q}$ be the wavelength routed from input $p$ to output $q$. Changing the input position, for instance to $p-p'$, will route the same $\lambda_{p,q}$ to ouput $q+q'$, with $p'=q'$ provided the positions of the input/outputs corresponds to the same wavelength displacement given by the derivative of Eq. (\ref{eq:theta}) (see \cite{smit}). Fig.~\ref{fig:srawg} shows a layout for N inputs and M outputs, accounting for these routing properties. The central input waveguide $p=0$ is placed a distance to the left from the center of the slab. Therefore, the central output waveguide $q=0$ needs to be placed the same distance to the right from the center. 

\subsubsection{Flat spectral response} \label{subsubsec:flat}
\begin{figure}
	\centering
	\includegraphics[width=0.6\textwidth]{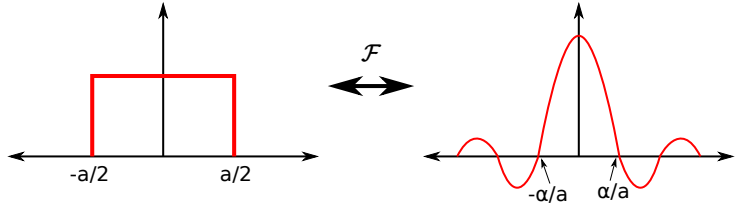}
	\caption{Rectangular and sinc functions.}
	\label{fig:rect_sinc}
\end{figure}

There are different techniques to flatten the spectral response of an AWG, amongst them the use of parabolic waveguide horns \cite{okamoto_horn}, MMIs \cite{pascual_mmi} and interferometers \cite{doerr_mzi} at the input/output waveguides. Other technique proposes the modification of the amplitude and phase in the AWs to obtain a sinc field profile \cite{okamoto}. The latter builds upon the signal theory duality between fields at both sides of the slab coupler, through the (spatial)  Fourier transform. To obtain a box like field pattern at the ouput side of the slab coupler, through the diffracted (the Fourier transform) field, a sinc distribution is required in the AWs \cite{okamoto_book, okamoto}. As mentioned in the introduction, the R-AWG concept proposed in this paper allows for the modification of the phase front by means of the phase shifters, while the amplitude can be adjusted by means of the SLRs. Recall the Fourier transform of a $\Pi$ function is:
\begin{equation}
	\mathcal{F} \left. \left\lbrace \prod \left( \frac{x}{A} \right) \right\rbrace \right|_{u=\frac{y}{\alpha}} = A \textrm{sinc} \left( A \frac{y}{\alpha} \right)
\end{equation}
where $A$ is the rectangular width, and $x$, $y$ the spatial variables. Therefore, the field at the plane $x_{2}$ will be modified to adjust it to a sinc function as described in \cite{okamoto_book}. In our formulation the adjustment can be incorporated by the terms in Eq. (\ref{eq:f2}), to be precise $2j \sqrt{\left(1-k_{r} \right) k_{r}} e^{-j2\psi_{PS,r} \left( \nu \right)} e^{-j \beta l_{SLR,r} \left( \nu \right)}$, from which the following amplitude and phase conditions are derived to turn the input far field Gaussian profile $B_{i} \left( x_{2} \right)$ into a sinc function. Hence, the amplitude condition is written as follows:
\begin{equation} \label{eq:amp_condition}
	B_{i} \left( r d_{w} \right) \sqrt{\left(1-k_{r} \right) k_{r}} = \left| \mathrm{sinc} \left( a \frac{rd_{\omega}}{\alpha} \right) \right|
\end{equation}
where $a$ will be the obtained rectangular function width when using the Fourier transform, as is detailed below. In addition, the following phase condition is required to turn the all positive values from the input far field Gaussian into negative ($\pi$ shift) where needed:
\begin{equation} \label{eq:phase_condition}
	2 \psi_{PS,r} \left( \nu \right) + \beta l_{SLR,r} = \left\{ \begin{array}{lll}
        0, & \mbox{if $\frac{2n\alpha}{a} \leq \left| rd_{\omega} \right| \leq \frac{\left(2n+1\right)\alpha}{a}$} & \mbox{with n=0,1,...}\\
        \pi, & \mbox{otherwise} \end{array} \right. 
\end{equation}
Under these conditions, the sinc field profile for the AWs can be introduced in Eq. (\ref{eq:f2}) for $x_{2}$, resulting in:
\begin{equation} \label{eq:f2_sinc}
	f_{2} \left( x_{2}, \nu \right) = j \sqrt[4]{2\pi \omega_{g}^2} \left[ \prod \left( \frac{x_{2}}{Nd_{\omega}} \right) B_{i} \left( x_{2} \right) \phi' \left( x_{2}, \nu \right) \sum_{r=-\infty}^{+\infty} \delta \left( x_{2} - rd_{\omega} \right) \mathrm{sinc} \left( a \frac{rd_{\omega}}{\alpha} \right) \right] \otimes b_{g} \left( x_{2} \right) 
\end{equation}
Finally, the field at the plane $x_{3}$ calculated through the spatial Fourier transform of Eq. (\ref{eq:f2_sinc}) is given by:
\begin{equation}
	f_{3,p} \left( x_{3}, \nu \right) = j \sqrt[4]{2\pi \frac{\omega_{g}^{2}}{\alpha^{2}}} B_{g} \left( x_{3} \right) \psi' \left( \nu \right) \sum_{r=-\infty}^{+\infty} f'_{M} \left( x_{3} + pd_{i}-r\frac{\alpha}{d_{\omega}} + \frac{\nu}{\gamma} \right)
\end{equation}
where $f'_{M} \left( x_{3} \right)$ is in this case:
\begin{equation} \label{eq:fmp}
	f'_{M} \left( x_{3} \right) = \mathrm{sinc} \left( N d_{\omega} \frac{x_{3}}{\alpha} \right) \otimes \frac{\alpha}{a} \prod \left( \frac{x_{3}}{a} \right)
\end{equation}
An important aspect for the design of flat-top AWGs following this technique is the number of AWs sinc field distribution zeros. Eq. (\ref{eq:fmp}) describes the field shape at the output plane, being the second term in the equation the obtained rectangular profile from the sinc field distribution in the AWs. Fig.~\ref{fig:rect_sinc} illustrates the relation between the positions of the sinc zeros and the width of the obtained dual rectangular function.
In general, there is a trade-off between the desired channel flatness and the acceptable channel width increase, which is inversely proportional to the number of zeros of the sinc field distribution in the AWs. Furthermore, the use of a more ``compressed'' sinc will reduce the amplitude of the obtained dual rectangular function, i.e. the AWG will have a flatter response, but with more peak insertion loss.


\section{\label{sec:sims}Simulations}
In this section, design cases with transfer function simulation using the equations above (similar to previously validated models \cite{munoz_validation, leijtens}) is presented. Despite SLRs can be implemented in nearly all integration technologies \cite{munoz_icton2013}, the footprint advantage will be so only in those where the confinement in optical waveguides is strong, i.e. the bend radius can be small. Amongst the different waveguide technologies available, the smallest bend radius is for Silicon-on-Insulator (can be less than 5 $\mu$m). Therefore the case provided in the following is for SOI technology, with a 220-nm-thick Si guiding layer on a SiO$_{2}$ substrate with no cladding. The effective indexes, calculated using a commercial software, are 2.67 in the arrayed waveguides ($n_{c}$) -waveguide with 0.8 $\mu$m to minimize phase errors, see \cite{bogaerts}- and 2.83 in the slab coupler ($n_{s}$). The R-AWG parameters are the following: the center wavelength is 1550 nm, using 6 channels with a spacing of 1.6~nm and a FSR of 19.2~nm. The calculated focal length is 217.37~$\mu$m, the incremental length between AWs is 31.38~$\mu$m and the number of AWs is 57. The bend radius was set to 5 $\mu$m, and the SLR loop length was set to a circumference of that radius, 31.4~$\mu$m. The design makes unse of a single input waveguide $i_0$ placed at the center position of the slab coupler input/output side. Consequently, the output waveguides are divided in tow halves, each at one different side of the input waveguide. This will result into a wavelength displacement of half channel (0.8 nm) in the output spectra, with respect to the design wavelength \cite{takahashi}. The motivation to use this special input/output configuration is given by the fact a Rowland mounting is used as input/output plane. In the case the input waveguide is displaced from the center, additional fine tuning techniques are required to compensate for the non-uniformities that arise, for example the modification of the angle and position for the waveguides as described in \cite{beelen_patent}.

\subsection{Gaussian and flatenned responses}
\begin{figure}
	\centering
	\subfloat[]{
		\label{fig:rawg_sinsinc_aw}
		\includegraphics[width=0.48\textwidth]{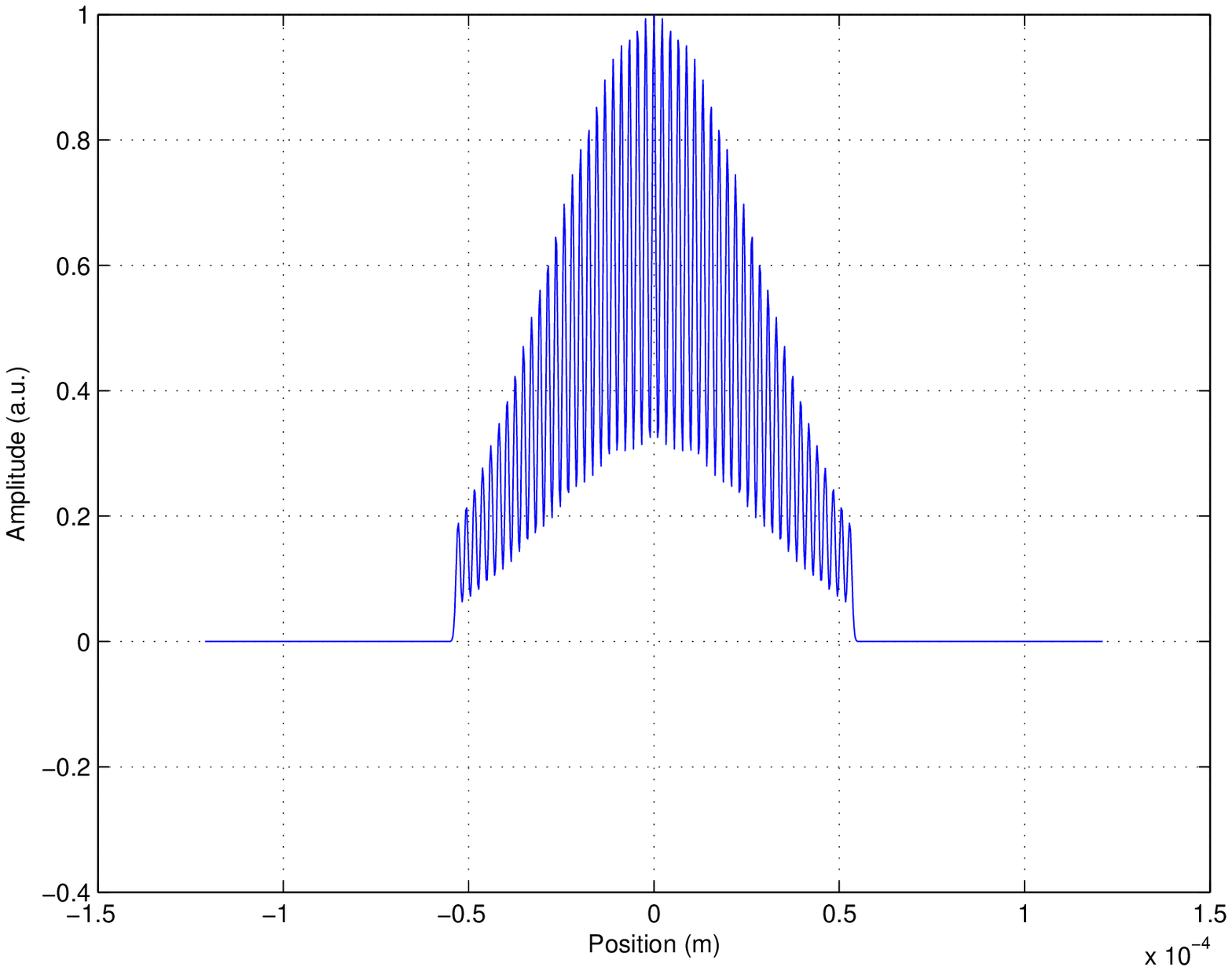}}
	\subfloat[]{
		\label{fig:rawg_sinsinc_t}
		\includegraphics[width=0.48\textwidth]{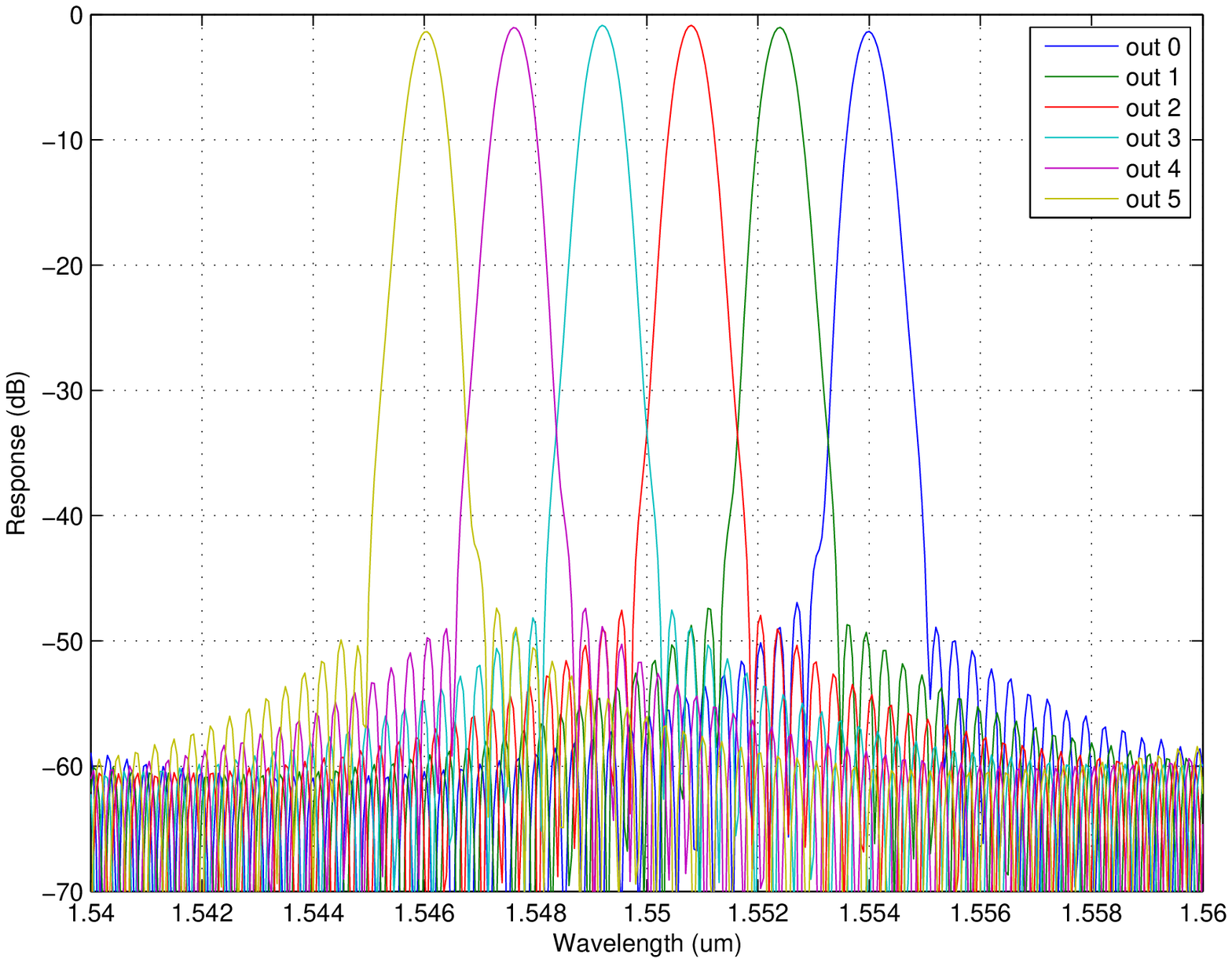}}
	\caption{Gaussian R-AWG simulation with 1 input and 6 outputs. (a) Field at the arrayed waveguides. (b) Transfer function from $i_{0}$ to the output waveguides.}
	\label{fig:rawg_7io_sim}
\end{figure}

The first set of simulations is for a Gaussian device as described in Section \ref{subsec:gaussian}. The response was calculated, without phase shifters and assuming all SLRs have ideally identical coupling constant $k=0.5$, i.e. total reflection. Fig.~\ref{fig:rawg_sinsinc_aw} shows the Gaussian field distribution at the plane $x_{2}$ obtained as the summation of all the AW contributions through Eq. (\ref{eq:f2_sinsinc}). The corresponding end-to-end transfer function for this R-AWG is depicted in Fig.~\ref{fig:rawg_sinsinc_t}. 

Note the simulations show losses of approximately 1~dB for the central channel and lower than 2~dB for the side channels. We did not include the propagation loss in the waveguides for SOI (typically around 4 dB/cm) and other detrimental effects as fabrication imperfections. The actual peak insertion loss of a regular SOI AWG can be as low as 4-5 dB. From the simulation the 1-dB, 3-dB and 20-dB bandwidths are 0.37~nm, 0.65~nm and 1.64~nm respectively.

\begin{figure}
   \centering
   \subfloat[]{
        \label{fig:field_aw_sinc35}
        \includegraphics[width=0.48\textwidth]{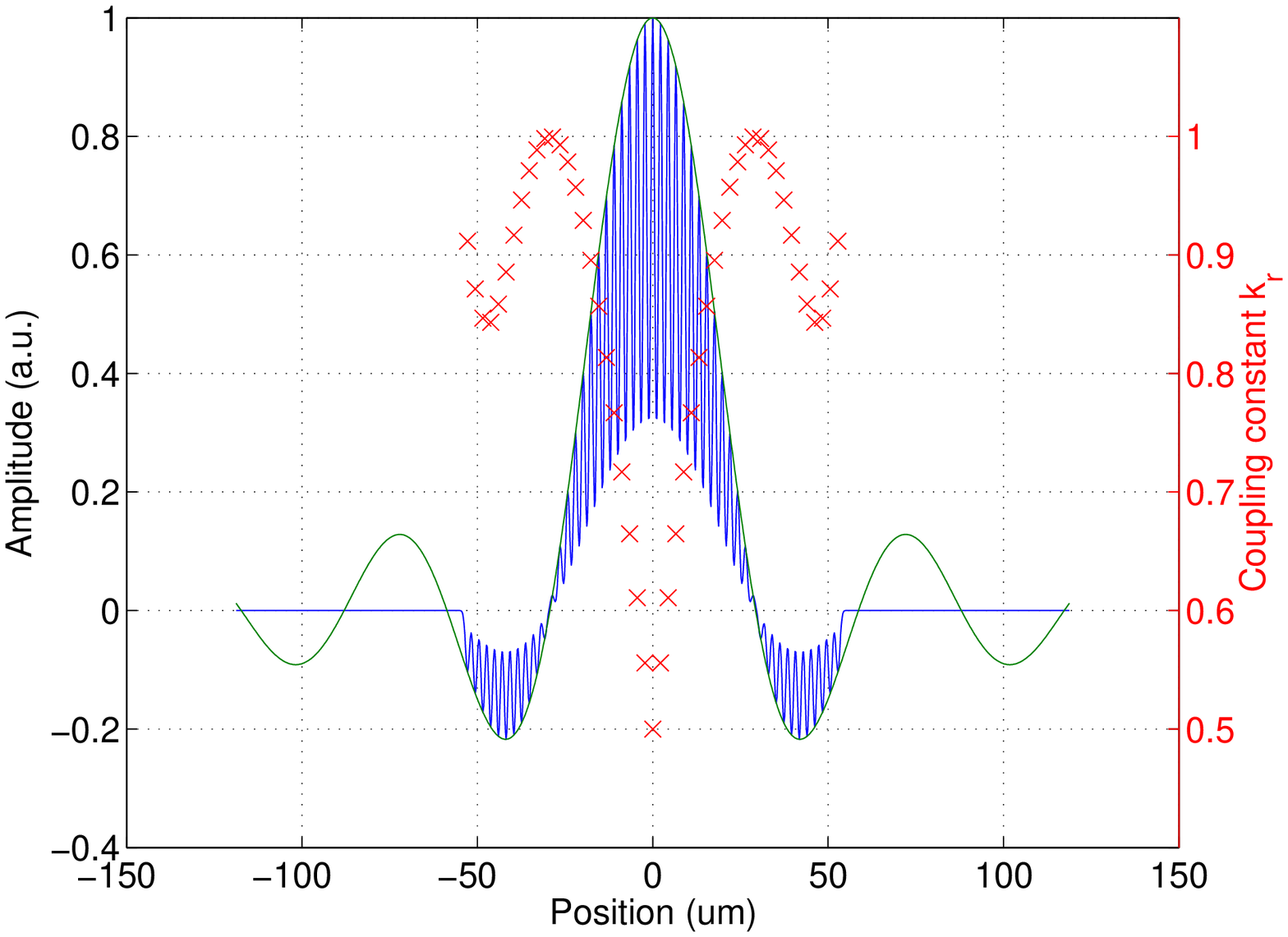}}
   \subfloat[]{
        \label{fig:t_output_sinc35}
        \includegraphics[width=0.48\textwidth]{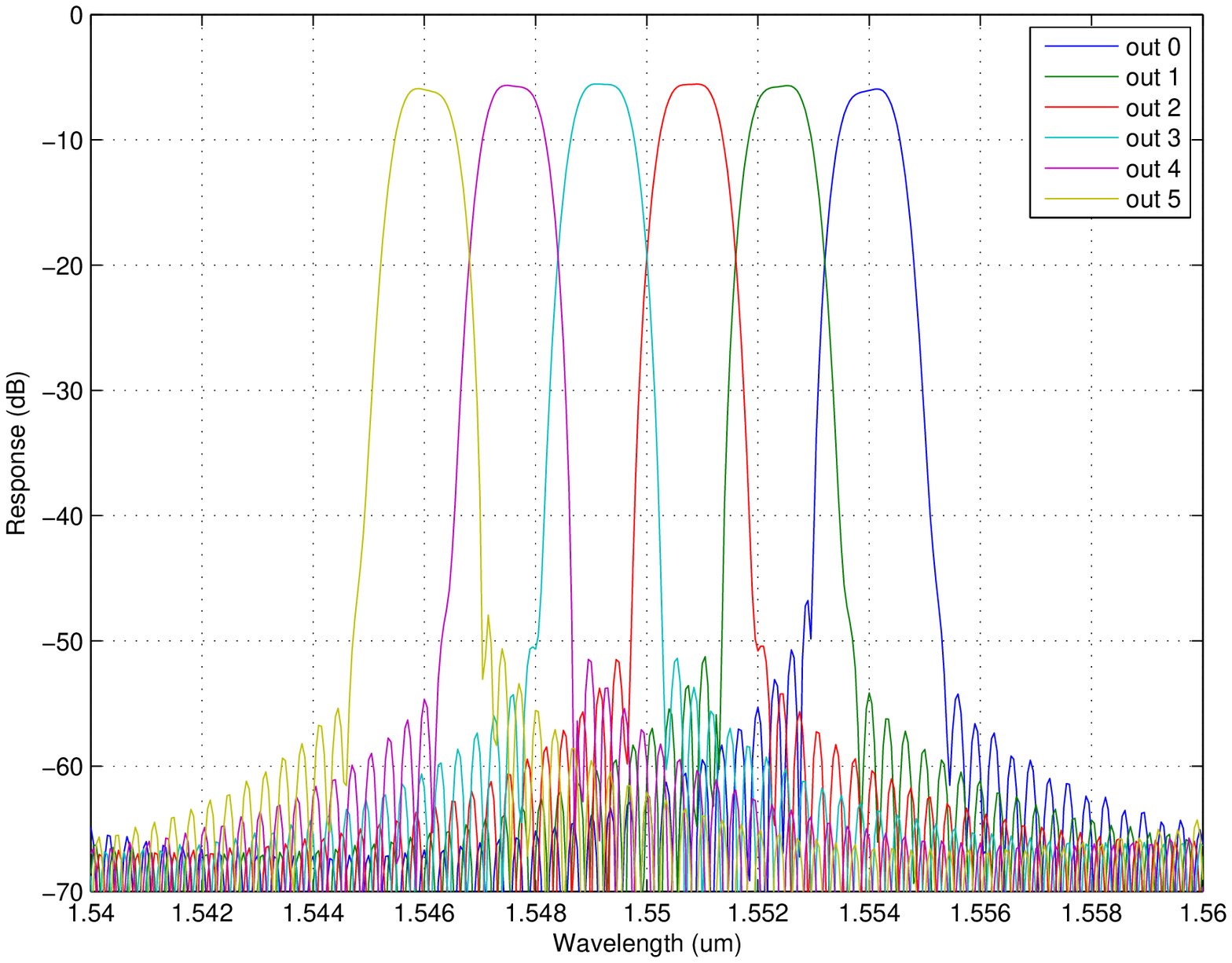}}
   \caption{Flat-top R-AWG using a sinc field distribution at the arrayed waveguides. (a) Field at the arrayed waveguides (blue), the sinc profile applied (green) and SLR coupling constant  $k_{r}$ in each arm of the array (red crosses). (b) Transfer function from $i_0$ to the output waveguides. (Both for a sinc distribution with parameter a=3.5$\mu$m).}
   \label{fig:t_output}
\end{figure}

The second set of simulations employ the same physical parameters, but for a R-AWG with sinc field distribution in the AWs, resulting in a flattened spectral response. A sinc profile with a~=~3.5~$\mu$m is incorporated. From this distribution, the required coupling constant $k_{r}$ for each SLR is calculated using Eq. (\ref{eq:amp_condition}). Note the use of a different coupler in each AW may introduce a different phase shift in each arm \cite{besse}, as mentioned in the introduction. For simplicity, this phase shift has not been included in Eq. (\ref{eq:phase_condition}) since it can be compensated through the phase shifters. Fig~\ref{fig:field_aw_sinc35} shows the field distribution at the plane $x_{2}$, being this field the summation of all the AW contributions in blue trace. On the same figure, the sinc function applied is shown in green line. Moreover, the secondary axis shows in red crosses the required coupling constant for each SLR to obtain the sinc profile. 

As described in the previous section, to obtain a wider rectangular function, a more compressed sinc function at the AWs is required. However, widening comes at the expense of increased channel insertion loss. This can also be understood by comparing Fig.~\ref{fig:rawg_sinsinc_aw} and Fig~\ref{fig:field_aw_sinc35}, from which is clear the sinc field distribution is attained in part by modifying the amplitude of the original Gaussian field distribution, with partial reflectors, i.e. some signal is lost. The transfer function for the flat-top R-AWG is shown in Fig.~\ref{fig:t_output_sinc35}. The flat spectral response and increased insertion losses are clearly noticeable. The obtained losses in this case are 5.6~dB and 6.7~dB for the central and side channels respectively. The bandwidths at the points of interest are in this case 1.01~nm, 1.36~nm and 2.37~nm, for 1-dB, 3-dB and 20-dB fall from the channel center. As expected, an increase in the channel bandwidth is attained at the expense of more insertion losses.

\begin{figure}
\centering
\subfloat[]{
\label{fig:field_all}
\includegraphics[width=0.5\textwidth]{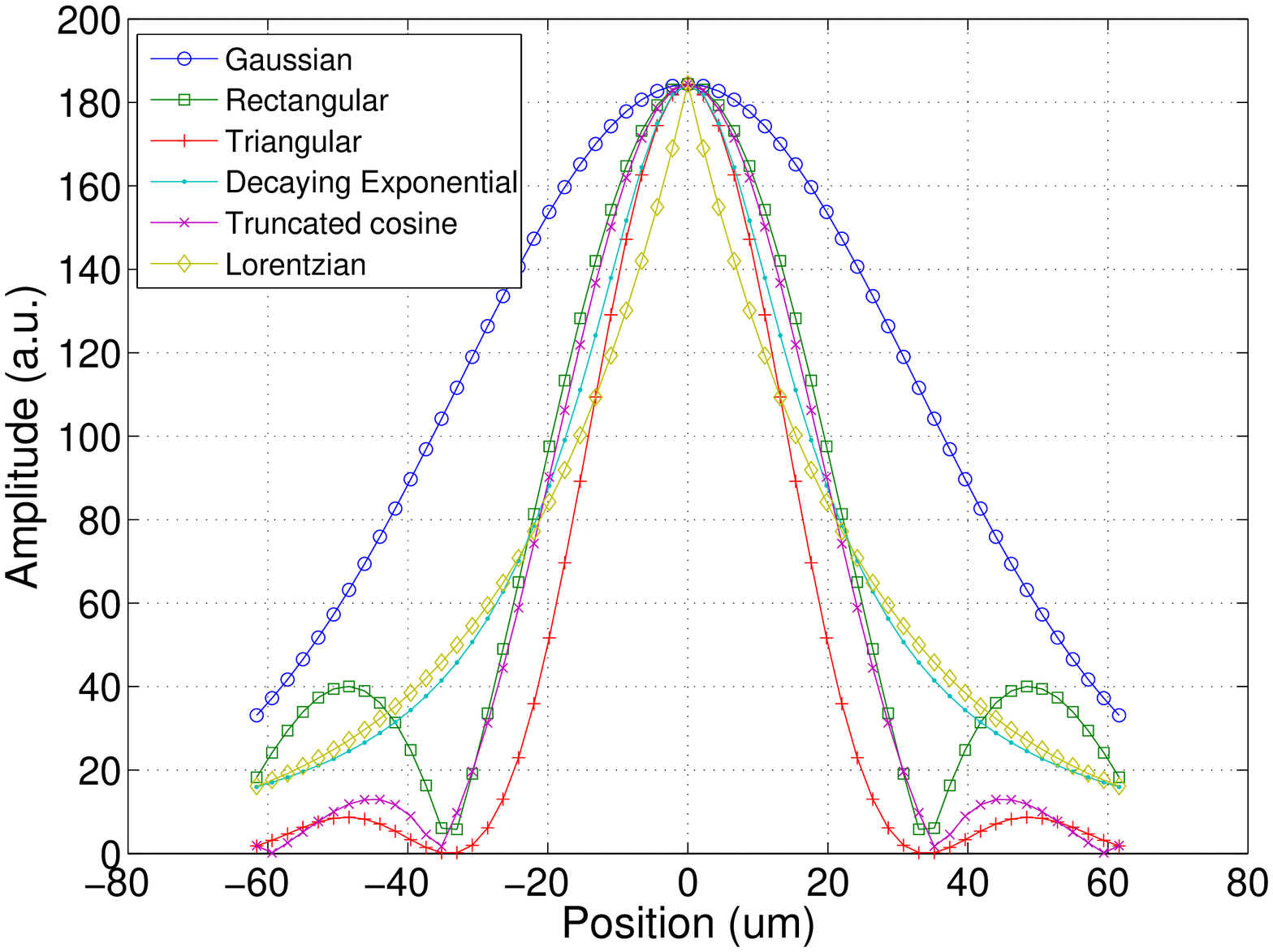}}
\subfloat[]{
\label{fig:k_all}
\includegraphics[width=0.5\textwidth]{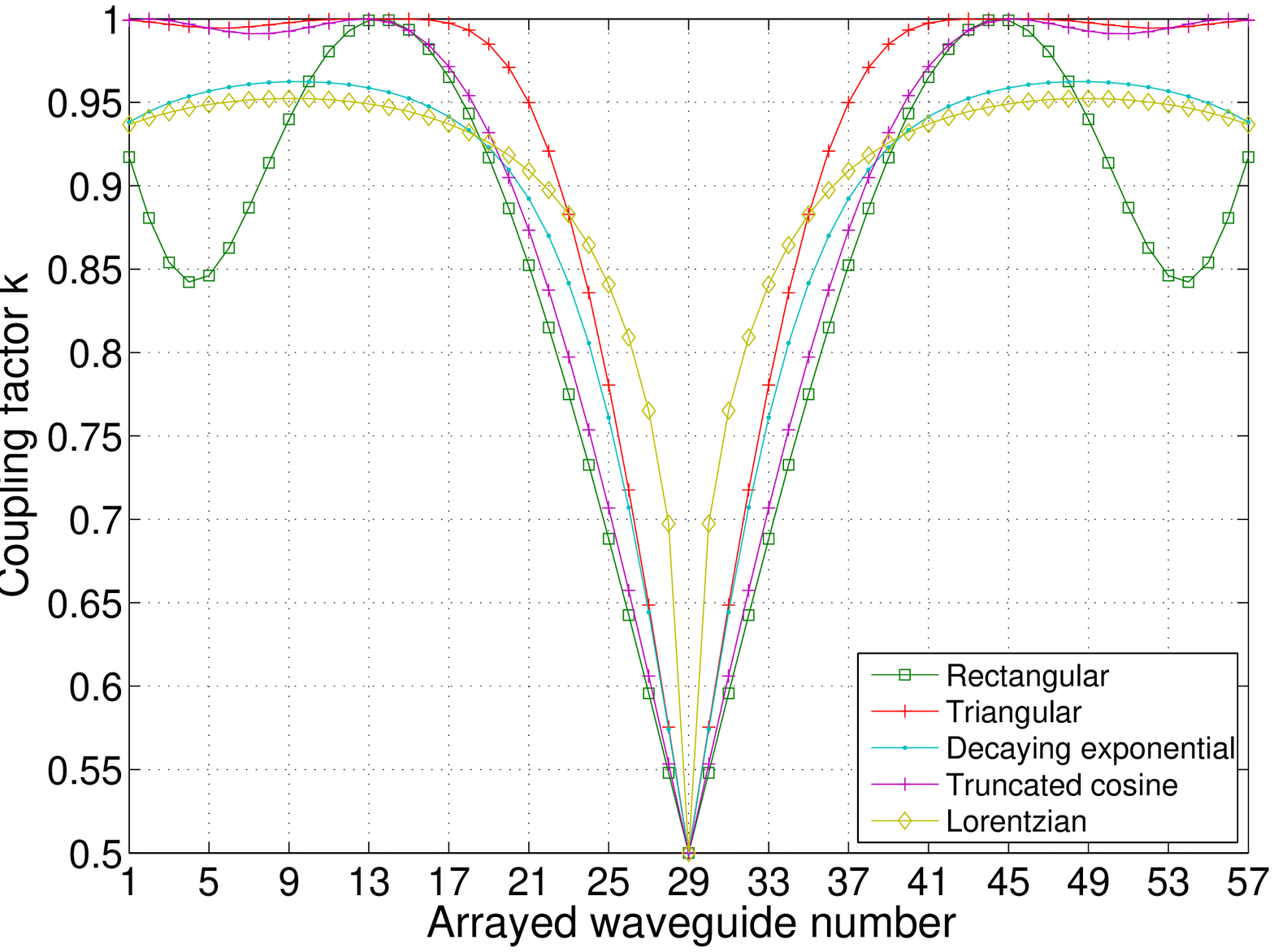}}
\caption{(a) Field at the AWs when different profiles have been applied. (b) Required coupling factor parameter at the SLRs when applying each profile.}
\label{fig:field_k}
\end{figure}

\subsection{Arbitrary spectral responses}
\begin{figure}
\centering
\includegraphics[width=0.8\textwidth]{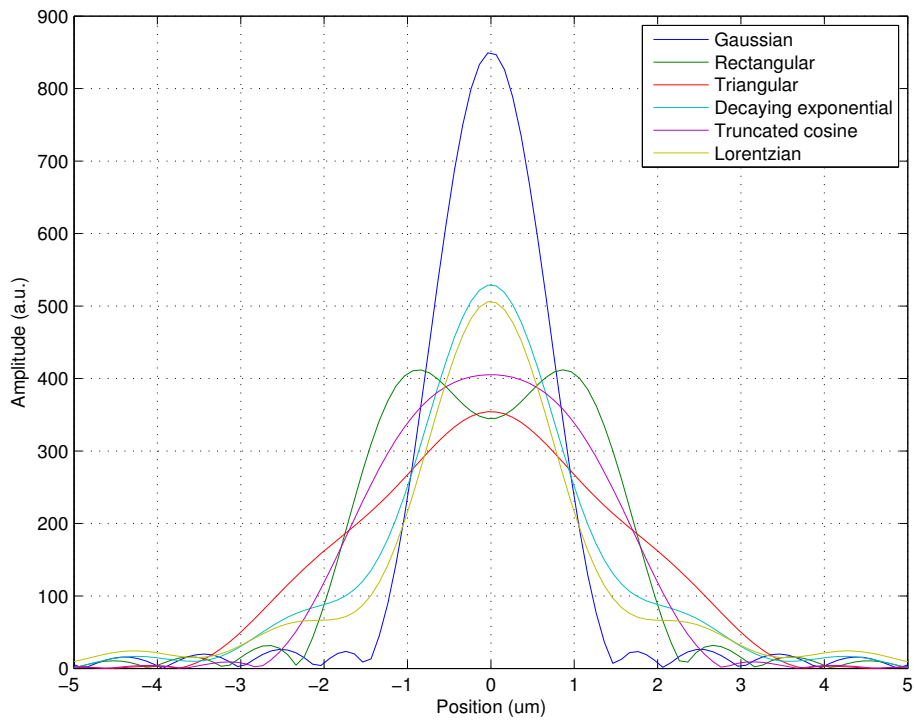}
\caption{Field focused at the output plane when using the central wavelength ($\lambda_{0}$) for each different profile applied at the AWs.}
\label{fig:fosN_all}
\end{figure}

Although the semi-analytical model in previous sections was only derived for the Gaussian and flattened response cases, in principle it is possible to apply any desired field distribution for the AWs, which will result in different spectral responses. In this subsection we present several field distributions and corresonding spectral responses, which we take from well-known Fourier transform pairs. To be precise, we targetted triangular, decaying exponential, truncated cosine and Lorentzian spectral responses. Fig.~\ref{fig:field_k}\subref{fig:field_all} shows the required AWs field distributions, i.e. at plane $x_{2}$. Note the legend labels are for the target transform pair, not the actual function employed in the field distribution for the AWs. Detailed expressions can be find elsewhere, as for instance in \cite{ft_pairs}.

Similar to the previously shown case for the flattened response (AWs sinc distribution), the required coupling factors $k$ to be applied in each AW are shown in Fig.~\ref{fig:field_k}\subref{fig:k_all}. Note there is no plotted value for the Gaussian case, since all the SLRs are use $k=0.5$ for full reflection. An additional important remark is how the regular Gaussian field distribution ingoing to the array is transformed into the targeted one. In principle some field distributions, can have amplitudes higher than those of the Gaussian for some of the waveguides in the array. This would required amplification, which is not contemplated with the proposed SLR-based layout. Therefore, the targeted profile needs to be inscribed under the starting Gaussian profile. Hence, the amplitude in each AW needs to be reduced to inscribe the profile inside the Gaussian, at the cost of more insertion losses. This is the result in Fig.~\ref{fig:field_k}\subref{fig:k_all}, where all the field profiles at the AWs have amplitude levels below the starting Gaussian distribution.

\begin{figure}
\centering
\includegraphics[height=0.4\textheight]{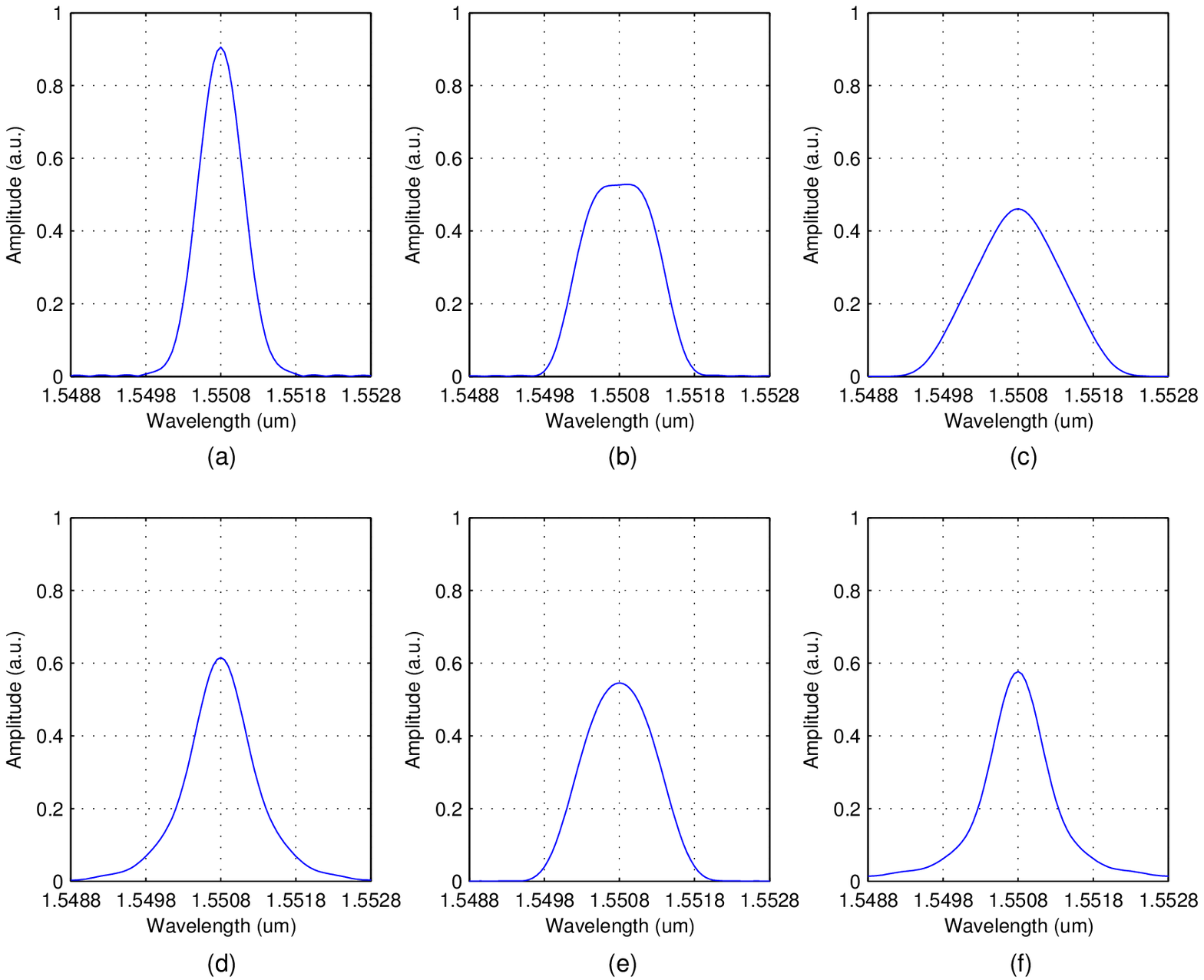}
\caption{Transfer function (linear) in one output waveguide for each different profile applied: (a) Gaussian, (b) rectangular, (c) triangular, (d) decaying exponential, (e) truncated cosine and (f) Lorentzian functions.}
\label{fig:t2_all_lineal}
\end{figure}
\begin{figure}
\centering
\includegraphics[height=0.4\textheight]{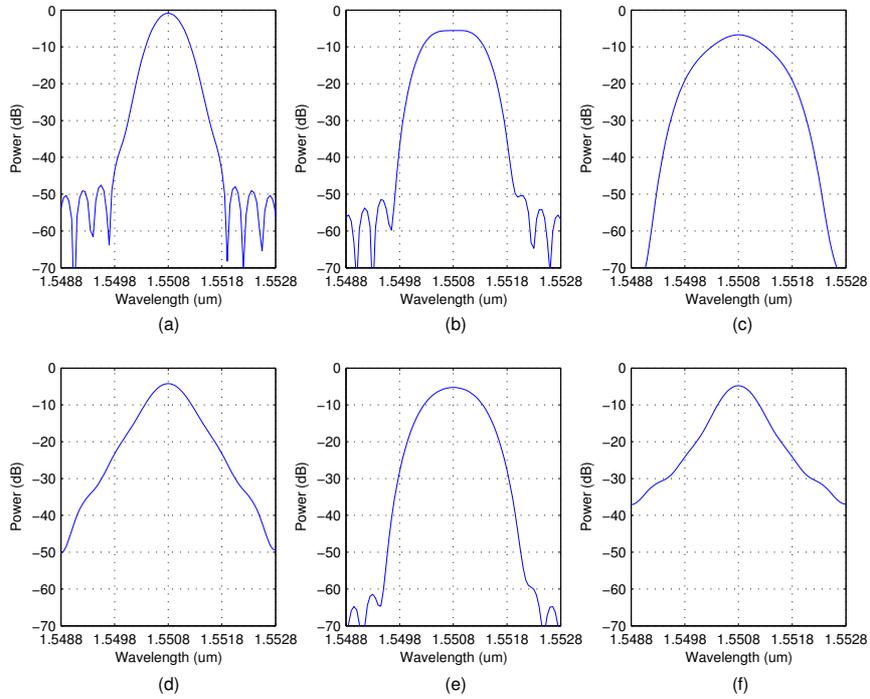}
\caption{Transfer function (logarithmic) in one output waveguide for each different profile applied: (a) Gaussian, (b) rectangular, (c) triangular, (d) decaying exponential, (e) truncated cosine and (f) Lorentzian functions.}
\label{fig:t2_all}
\end{figure}

The flat spectral response case developed in Section \ref{subsubsec:flat} shows the shape of the field at the output plane is given by Eq.~(\ref{eq:fmp}). For all the targeted spectral responses, this far field at the output plane is plotted in Fig.~\ref{fig:fosN_all} for $\lambda_{0}$. The far field is not exactly the Fourier transform pair of the AWs field distribution in Fig.~\ref{fig:field_k}\subref{fig:field_all}. As expressed in  Eq.~(\ref{eq:fmp}), the field profiles at $x_2$ have a finite extensions (i.e. a finite number of AWs is employed), therefore the field profiles are truncated and the far field is in fact the convolution between a sinc function (Fourier transform of a truncation function in the array) and the Fourier transform of the profile applied at the AWs. From all the curves in Fig.~\ref{fig:fosN_all} the triangular function case (red line) is the most suitable to understand this fact. Ideally, for an infinity (unpractical) number of AWs, one would expect a perfect (sharp) triangular shape, but in practice the truncation by a finite number of waveguides results in some smoothing in the curves. This is similar to the well-know problem of function approximation with Fourier series using a finite number of terms.

In addition to this intrinsic smoothing, the corresponding end-to-end transfer functions for the R-AWGs involves the calculation of the convolution integral between the (already smoothed) far field at the output plane and the mode at the output waveguide, as described by Eq. (\ref{eq:tq}). The transfer functions are depicted in Fig.~\ref{fig:t2_all_lineal} and Fig.~\ref{fig:t2_all}, using linear and logarithmic units, for one output waveguide, in this case o$_{3}$ placed at a distance 2.24~$\mu$m from the slab center. 

\section{\label{sec:conclusion}Conclusion and outlook}
This paper proposes a novel type of reflective Arrayed Waveguide Grating, that makes use of a configuration based on phase shifters and Sagnac Loop Reflectors, build with an optical coupler with looped back waveguides. The layout enables the control of the field amplitude and phase per arm in the array, with the combination of phase shifters and SLRs whose reflectivity is set through the coupling constant of the optical coupler. A theoretical model for the analysis and design of the device has been provided, both for the cases of Gaussian and flattened response, the latter achieved by adjusting the AW field distribution to a sinc function by means of the SLRs and phase shifters. The model was used to design and simulate Silicon-on-Insulator implementations using typical waveguide cross-sections for the technology. This was presented for the Gaussian and flattened spectral response cases, as well as for different field distributions in the AWG arms that result in principle with an arbitrarily customizable spectral response.

We believe profiles in the AWs can be found to pre-equalize the intrinsic spectral response smoothings described, in order to obtain closer to target spectral responses. The processing of photonic signals in the wavelength domain, i.e. multi-wavelength spectral filtering/shapping, is just one of the possible applications of this versatile and novel R-AWG layout. We envisage more applications for which the field distribution in the AWs is determinant, as the use of AWGs for pulse rate multiplication, where the envelope of the train of pulses generated by the AWGs is directly dictated by the field distribution in the arms \cite{leaird}.

\section*{Acknowledgment}
The authors acknowledge financial support by the Spanish MICINN project TEC2010-21337, acronym ATOMIC, the MINECO project TEC2013-42332-P, acronym PIC4ESP, project FEDER UPVOV 10-3E-492 and project FEDER UPVOV 08-3E-008. B. Gargallo acknowledges financial support through FPI grant BES-2011-046100. The authors thank J.S. Fandi\~no for helpfull discussions.


\end{document}